# Unexpected Xe cations and superconductivity in Y-Xe compounds under pressure

Dawei Zhou [a], Dominik Szczęśniak[b], Jiahui Yu [c], Chunying Pu [a**], Xin Tang [d]

[a]*College of Physics and Electronic Engineering, Nanyang Normal University, Nanyang 473061, China*

[b]*Institute of Physics, Jan Długosz University in Częstochowa, Ave. Armii Krajowej 13/15, 42-200 Częstochowa, Poland*

[c]*School of Electronic and Electrical Engineering, Nanyang Institute of Technology, Nanyang 473004, China*

[d] *College of Material Science and Engineering, Guilin University of Technology, Guilin 541004, China*

**Abstract**

The metal-based noble gas compounds exhibit interesting behavior of electronic valence states under pressure. For example, Xe upon compression can gain electrons from the alkali metal, or lose electrons unexpectedly to Fe and Ni, toward formation of stable metal compounds. In addition, the $Na_2He$ is not even stabilized by the local chemical bonds but via the long-range Coulomb interactions. Herein, by using the first-principles calculations and the unbiased structure searching techniques, we uncover that the transition metal Y is able to react with Xe above 60 GPa within various Y-Xe stochiometries, namely the YXe, $YXe_2$, $YXe_3$ and $Y_3Xe$ structures. Surprisingly, it is found that all the resulting compounds are intermetallic and Xe atoms are positively charged. We also argue that the pressure-induced changes of the energy orbital filling are responsible for the electron transfer from Xe to Y. Meanwhile, the Peierls-like mechanism is found to stabilize the energetically most favorable YXe-*Pbam* phase. Furthermore, the predicted YXe-*Pbam*, YXe-*Pnnm*, and $YXe_3$-*I*4/*mcm* phases are discovered to be a phonon-mediated superconductors under pressure, with the critical superconducting temperatures in the range of approximately

** Corresponding author email: puchunying@126.com

3-4K, 7-10K, and 5-6K, respectively. In summary, our work promotes further understanding of the crystal structures and electronic properties of the metal-based noble gas compounds.

# 1. Introduction

The noble gas (NG) elements (e.g. He, Ne, Ar, Kr and Xe) are historically known as inert elements due to their low level of reactivity arising from the completely filled outermost electron shells (the closed subshell configuration). However, since the first experimentally proved synthesis of the metal-based NG compound by Bartlett[1], many other NG compounds were found to exist [2-6]. Specifically, most of the early NG compounds were synthesized by oxidizing the nobel gas elements with the highly electronegative elements such as fluorine[7-8] and oxygen[9-12]. Yet later, it was also found that the high-pressure compression constitutes another suitable and promising route for the synthesis of NG compounds. For example, the compression of noble gases and some molecular species leads to the formation of multiple van der Waals solids at relatively low pressure, such as: $HeN_4$[13], He-$H_2O$[14], $Ar(H_2)_2$[15], $ArO_2$, $Ar(O_2)$[16], $Kr(H_2)_4$[17], $Xe(H_2)_8$[18], Xe-$N_2$[19], Xe-$CH_4$[20], Xe-$H_2O$[21] and $Xe(O_2)_2$[16]. Moreover, pressure can also enhance the oxidation of some elements and results in the formation of noble gas compounds such as Xe-oxides[22], and Xe-suboxides[23], as well as promote the formation of covalent bonds as in the $Xe_2F$ compounds[24]. Another striking pressure effect in the NG and metal compounds is the unexpected behavior of electronic valence states. For example, Xe can gain electrons from some metals (the alkali and the alkaline earth metals) and form stable metal-compounds such as the Li-Xe, Cs-Xe and Mg-Xe[25-27]. On the contrary, Xe is also found to be a reducer, being oxidized by Fe and Ni under the pressure of Earth's core[28-29], which is believed to be the answer to the missing Xe on Earth. Furthermore, very recently the mixtures of sodium (as well as its oxide) with helium were reported to be stable under pressure[30-31]. Therein, helium shows new chemistry under pressure, and does not lose electrons nor form any chemical bonds in the $Na_2He$ and $Na_2OHe$ compounds, suggesting the long-range Coulomb interactions to be responsible for the formation of these compounds[31]. In this regard, the electronic valence state of metal in the pressure-induced metal-based NG compounds exhibits some degree of uncertainty, providing a very broad scenario which still awaits further investigation.

In this context, the compressed compounds of NGs and transition metals appear as a particularly interesting systems. Of special attention to us are the Xe-Fe(Ni)[28-29] compounds, where Fe(Ni) does not behave as a usual metal under pressure and surprisingly plays a role of an oxidant (in contrast to Li, Cs and Mg). Nonetheless, we note that the transition metal family has a relatively large number of members with similar properties to Fe(Ni). Therefore, two natural questions concerning the metal-based NG compounds arise. First, can all the other transition metals form stable NG compounds under pressure just like Fe(Ni)? Second, does such transition metals gain or lost electrons when reacting with NG under pressure? To at least partially address above questions, herein we explore the potential Y-Xe compounds under pressure. In relation to the Fe-family elements, the position of Y in the periodic table of elements is distant. Moreover, at ambient conditions, the electronegativity of Y in the Pauling scale is 1.22, which is similar with that of Mg (1.31), but significantly lower than in the case of Fe/Ni (1.83/1.91). On the other hand, Xe is of particular attention due to its rich chemistry when combined with the iron-family elements and its role of a well-established NG host for other metals, as described above. Hence the proposed investigations on the Y-Xe compounds appear as a representative example which may have potential for exploring novel chemistry of the metal-based NG compounds under pressure.

In the present work, motivated by the above reasoning, we extensively explore the high-pressure phase diagram of Y-Xe systems up to 200 GPa by using the CALYPSO structural searching method[32-34]. Such method for crystal structure prediction is chosen due to its well-proven accuracy for various systems[35-39]. In its framework we attempt to uncover that when mixing Xe with Y under high pressure, the thermodynamically stable compounds with various stoichiometries can be formed. Next, the predicted structures are comprehensively analyzed with respect to their structural and electronic properties, toward further examination of the potential non-conventional behavior.

# 2. Computational Methods

In the present paper, the structure search for the Y-Xe systems is performed by using the particle-swarm optimization method, as implemented in the CALYPSO code[32-34]. The computations are carried out from one up to four formula units per simulation cell at 50, 100, 150, and 200 GPa, respectively. All the structural relaxations and the electronic calculations are conducted in the framework of density functional theory within the projector augmented wave (PAW) method[40], by using the Vienna ab initio simulation package (VASP).[41-42] Herein, the Y $4d^8 5s^2 5p^1$ electrons and the Xe $4d^{10} 5s^2 5p^6$ electrons are treated as valence. The exchange and correlation interactions are considered within the GGA[43] approach by adopting the Perdew-Burke-Ernzerhof functional[44]. To ensure that the total energy calculations are well converged, a cutoff energy of 700 eV and the Monkhorst-Pack[45] (MP) *k*-point mesh with a reciprocal space resolution of $2\pi\times0.025$ Å$^{-1}$ is assumed. The convergence for the energy and force is chosen as $10^{-6}$ eV and $10^{-3}$ eV/ Å, respectively.

Furthermore, the phonon dispersions and the electron-phonon coupling (EPC) calculations are performed in the framework of the linear response theory by employing the QUANTUM-ESPRESSO package[46]. The ultrasoft pseudopotentials[47] of Xe and Y are used with a kinetic energy cutoff of 60 Ry. A 3×5×8, 4×2×6, 4×4×4, 4×4×4, and 6×6×3 *q*-point mesh in the first Brillouin zone (BZ) is used within the EPC calculations for the Pbam-YXe, Pnnm-$YXe_3$, *I*4/*mcm* $YXe_3$, *I4/mmm* $YXe_2$, and *P4/mmm* $YXe_3$ stoichiometries, respectively. Accordingly for the listed phases, a MP grid of 6×10×16, 8×4×12, 8×8×8, 8×8×8, and 12×12×6 is assumed to ensure the *k*-point sampling convergence.

# 3. Results and discussion

To investigate the phase stability of the Y-Xe systems under pressure, we calculate the convex hull at the selected pressures on the basis of the Y-Xe formation enthalpies, as shown in Fig.1a. The formation enthalpies are estimated as a difference between the enthalpy of the predicted Y-Xe structures with elemental Y and Xe. The known hcp Xe[48], as well as the dfcc and *Fddd* Y[49-50] are chosen as the reference structures in their corresponding stable pressure ranges, respectively. In this context, the compounds on the convex hull are thermodynamically stable, while those above it are metastable. As shown in Fig.1a, the convex curve reveals that all stoichiometries are not stable against elemental dissociation below 50 GPa. However, at about 60 GPa, we find that YXe and $Y_3Xe$ stoichiometries become energetically stable. As pressure is increased, also the $YXe_2$ and $YXe_3$ are predicted to be stable against the decomposition at 100GPa and 150 GPa, respectively. We notice that among all found Y-Xe stoichiometries, YXe is the most thermodynamically stable phase, in the pressure range from 75 GPa to 200 GPa. For convenience, Fig.1b shows the pressure-phase diagram of all predicted YXe compounds, while their graphical representations are shown in Fig. 2. Moreover, the structural parameters of predicted compounds are listed in Table 1. In particular, the YXe up to 200 GPa is a triclinic structure with space group *Pbam*. On the other hand, the $Y_3Xe$ adopts an orthorhombic structure with space group *Pnnm* at 75 GPa, while it transforms to a tetragonal structure with space group *I*4/*mcm* at about 80 GPa. In what follows, the predicted $YXe_2$ adopts an *I*4/*mmm* space group and no further phase transition is found in the considered pressure range. To this end, for the $YXe_3$, the phase transition from P3m1 to *P*4/*mmm* occurs at about 78 GPa. As shown in the convex hull curves, although $YXe_3$ becomes a stable phase against elemental dissociation only above 150 GPa, the low-pressure *P*-3*m*1 phase may exists as a metastable phase.

Next, the electronic band structures, and the partial density of states (DOS) of the predicted Y-Xe compounds are investigated at selected pressure values, and the corresponding results are depicted in Fig.3. The analysis reveals that all Y-Xe compounds exhibit metallic character due to several bands crossing the Fermi level. From the partial DOS we can see that for all stable Y-Xe compounds the conducting states mainly derive from the Y-5d states in the vicinity of the Fermi level. However, the Y-s states also give moderate contribution around the Fermi level, especially for the $Y_3Xe$-*Pnnm*, $Y_3Xe$-*I*4/*mcm* and $YXe_2$-*I*4/*mmm* structures. It is not a big surprise since the s-d transformation of the Y solid metal under pressure has been reported in the previous studies[51,43]. Moreover, we notice that the Xe-5p orbitals contribute mostly to Xe states around the Fermi level for the $YXe_2$-*I*4/*mmm* structure, while for the other predicted structures the Xe-5d orbitals show the most significant contribution, especially for the YXe-*Pbam* and $YXe_3$- *P*4/*mmm* stochimetries.

As mentioned above, the metal-based Xe compounds exhibit unusually electronic valence states under pressure. To investigate the electron transfer and chemical bonding of Y-Xe compounds, we calculate the three dimensional charge difference density (CDC) maps of them, as shown in Fig.4. Although all the predicted structures are metallic, the covalent bonding can be found between Y atoms. It is also observed that Y atom gain electrons from Xe atom, and plays the role of an oxidant just like the Iron family metals. In fact, at ambient conditions, the electronegativity of Y is 1.22, according to the Pauling scale, which is lower than that of Xe (2.67). Therefore an electron transfer from Y to Xe atoms should be expected. However, in the predicted Y-Xe compounds, Xe is unusually positively charged and behaves as a reluctant rather than an oxidant. According to the previous work[52-53], pressure can dramatically affect the filling of energy orbitals and finally leads to the changes of element's electronegativity. For example, application of pressure results in the different behavior for a given metal. Specifically, Li[26, 54], Cs [27]and Mg[25] act as reducers, while Fe and Ni[28-29] act as oxidants in their corresponding metal-Xe compounds. To further reveal the origin of negatively charged Y atoms in the Y-Xe compounds, we use a method named "He matrix method"[55] toward the orbital distribution analysis for the Y and Xe atoms under pressure. Although this method is semi-quantitative to some degree, it successfully explains e.g. the reducibility of Li in Li-Xe compounds[54] under pressure. As depicted in Fig.5, we can see that the 5s, 5p and 5d orbitals of Xeon atoms are higher in energies than the 5d orbitals of Y atoms on the whole, an effect which promotes the electron transfer from the Xe to Y atoms.

Furthermore, we notice that the Xe atoms of the predicted structures show different coordination number. In particular, the coordination numbers of the Xe atom in the *Pbam*, *Pnnm*, and *I*4/*mmm* symmetry are four, eight and twelve, respectively, as shown in Fig.4. For the $YXe_3$-*I*4/*mmm* stoichiometry, we notice that four Y atoms and two Xe atoms compose a unique octahedron unit, the similar structure is also found for the $YXe_3$ *P*4/*mmm* phase. In addition to the unique octahedron unit, another body-centered like unit, with one Y and one Xe atom, is found for the *P*4/*mmm* phase. For both described types of unit cell, it can be clearly seen from the CDC of Y-Xe compounds that Xe atoms loses electron and transfers it to Y atoms.

Further, we investigate the structural characteristics of the energetically most favorable YXe *Pbam* phase, and found that if the Xe and Y atoms are arranged in line along the <110> direction, then the *Pbam* phase changes into the CsCl-like structure. In fact, the CsCl is also found to be a stable phase for the sibling Xe-Li and Mg-Xe systems under pressure[25-26]. Thus, it is argued that the *Pbam* phase can be regarded as a distorted CsCl structure. Of course, the distortions in YXe-*Pbam* will increase the Coulomb repulsion energy among the atoms, but if the distortions lead to the changes of the crystal electronic structure and the one-electron energy sum is reduced, then the distorted structure may become stable. The mentioned above phase transition mechanism is known as the Peierls mechanism [56]. To reveal the origin of distortions in the *Pbam* phase, we calculate the total and partial DOS of YXe in the *Pbam* and CsCl structures, which are shown in Fig.6. Just as expected, the distortions in the *Pbam* phase lead to the decrease of total DOS at the Fermi level. Further investigation of the partial DOS reveals that the decrease of total DOS at Fermi level mainly originates from the *d* states of Y and Xe atoms. Both the *d* states of Y and Xe atoms in the CsCl phase form a sharp peak at Fermi level, which is unfavorable energetically and finally results in a transition to the *Pbam* phase.

We also calculate the phonon curves and the superconducting critical temperature ($T_C$) of all the predicted YXe compounds. As shown in Fig.8, the phonon curves do not present any imaginary frequency mode, indicating that all the predicted compounds are dynamically stable. The $T_C$s of the metallic Xe−Y compounds at selected pressure are estimated using the Allen−Dynes modified McMillan equation[57] and the canonical value of $\mu$* equal to 0.1[58]. The corresponding results are listed in Table 2. In details, the $T_C$s of *Pbam*, *Pnnm* and *I*4/*mcm* phases under pressure are in the range of approximately 3-4K, 7-10K, 5-6K, respectively. It is noted that these values are higher than the $T_C$ of 0.04 K for Xe at 215 GPa, but lower than the maximum $T_C$ of 20 K for solid Y at 115 GPa. On the other hand, for the *I*4/*mmm* and *P*4/*mcm* phases, their $T_C$s are negligible. Moreover, the calculated electron-phonon coupling is 0.60 for YXe-*Pbam* (60 GPa), 1.08 for $Y_3$Xe-*Pnnm* (75 GPa) and 0.6 for $Y_3$Xe-*I*4/*mmm* (150 GPa), respectively, while the calculated average logarithmic frequencies, $\omega_{log}$, reach 181, 137 and 280 K, respectively. So the relatively strong electron-phonon coupling is found in the $Y_3$Xe-*Pnnm*. In contrary, for the Xe-rich compounds, namely the $YXe_2$-*I*4/*mmm* and $YXe_3$-*P*4/*mcm* phases, the electron-phonon coupling constants are estimated to be $\lambda$=0.29 (100 GPa), and 0.20 (150GPa), respectively, and lead to the nearly zero $T_C$s. The Eliashberg spectral functions together with the electron-phonon integrals $\lambda(\omega)$ are carried out to further investigate the superconducting properties of Y-Xe compounds as shown in Fig.7. Taken altogether, the vibrations of Y and Xe atoms are coupled strongly over the whole frequency area. From the magnitude of electron-phonon integrals $\lambda$, it can be observed that for the $Y_3$Xe-*Pnnm*, the comparatively low-frequency vibrations of Y and Xe atom below 3.5 THz contribute about 70% to the total $\lambda$. While for the other predicted stable Y-Xe compounds, the integral $\lambda$ increases linearly with frequency, indicating that all the vibrational modes have approximately equal contribution to the total $\lambda$.

# 4. Conclusions

In the present study, the phase stability of Y-Xe binary systems is examined in the pressure range from 0 to 200 GPa, by using the particle swarm optimization technique combined with the first principle calculations. It is shown that the first stable Y-Xe compounds (namely, YXe and $Y_3Xe$) can be formed above 60 GPa. However, upon further compression, new stoichiometries of $YXe_2$ and $YXe_3$ are also predicted to be energetically stable at 100 GPa and 150 GPa, respectively. In details, the unique octahedron unit and the body-centered like units are found in the Xe-rich compounds. The four, eight, and twelve coordination numbers of the Xe atom are identified in the Y-Xe compounds. The atoms distortions are predicted in the energetically most favorable *Pbam*-YXe phase, whereas the Peierls mechanism is found to stabilize the YXe-*Pbam* phase.

It is also suggested that all predicted Y-Xe compounds show metallic character. Unexpectedly in all cases, Y metal plays the role of an oxidant and the electrons are found to be transferred from Xe to Y atoms. We discover that pressure can greatly change the energy orbital filling of Y-Xe compounds, which results in the transfer of electrons from Xe to metal Y. Furthermore, except for the $YXe_2$-*I*4/*mm*m and $YXe_3$-*P*4*mmm* phases, all the other Y-Xe compounds are found to be phonon-mediated superconductors under pressure. In particular, the metal-superconductor transition temperatures of the *Pbam*, *Pnnm* and *I*4/*mcm* phases under pressure are in the range of approximately 3-4K, 7-10K, and 5-6K, respectively.

In what follows, it appears that the valence bond theory is facing great challenges under pressure, which application can greatly affect the energy orbital filling of solids and lead to the formation of new compounds with unusual behavior of the electronic valence states. In this context, the current theoretical study may serve as a representative example. Specifically, the formation of unique Y-Xe compounds indicates that the quantivalence of metal-based noble gas compounds under pressure remains uncertain, and suggests that the other metal oxidants, similar to Y, may also exist. In this sense, present work provides vital theoretical guidance for exploration and further experimental synthesis of the transition metal-based NG compounds.

## Acknowledgments

This work is supported in China by the National Natural Science Foundation of China (Grant Nos. 51501093, 41773057, U1304612, and U1404608), Science Technology Innovation Talents in Universities of Henan Province (No.16HASTIT047), Young Core Instructor Foundation of Henan Province (No. 2015GGJS-122).

**Figure and Table Captions**

**Fig.1** (a) Ground-state enthalpies of formation for various Y−Xe compounds, relative to the elemental constituents. The dashed lines connect the data points, while the solid lines denote the convex hull. (b) The phase diagram of Y–Xe system in the pressure range from 50 to 200 GPa.

**Fig.2** Stable phases for various Y-Xe compounds at selected pressures. (a) YXe-Pbam at 75 GPa (b)$Y_3Xe$-Pnnm at 75 GPa (c) $Y_3Xe$-*I*4/*mcm* at 100 GPa (d) $YXe_2$-*I*4/*mmm* at 100 GPa and (e) $YXe_3$-*P*4/*mmm* at 150 GPa. Green atoms depict Y, pink atoms present Xe.

**Fig.3** Calculated electronic band structures and partial DOS for various Y−Xe compounds. (a) YXe-Pbam at 75 GPa, (b)$Y_3Xe$-Pnnm at 75 GPa, (c) $Y_3Xe$-*I*4/*mcm* at 100 GPa (d) $YXe_2$-*I*4/*mmm* at 100 GPa and (e) $YXe_3$-*P*4/*mmm* at 150 GPa.

**Fig.4** Calculated charge difference density map of various Y-Xe compounds (a) YXe-Pbam at 75 GPa, (b)$Y_3Xe$-Pnnm at 75 GPa, (c) $Y_3Xe$- *I*4/*mcm* at 100 GPa (d) $YXe_2$-*I*4/*mmm* at 100 GPa and (e) $YXe_3$-*P*4/*mmm* at 150 GPa. Green atoms depict Y, pink atoms present Xe. Cyan and yellow colors represent losing and gaining electrons, respectively. The typical structure units of various Y-Xe compounds are illustrated at the left hand side of the corresponding structures.

**Fig.5** Simulated partial DOSs of Xe and Y atoms at 75 GPa using the "He matrix method". The pressure effect is simulated by placing elements in a face-centered-cubic (FCC) He matrix. An FCC 3×3×3 supercell with 108 atoms was used, in which one He is replaced by the tested atoms.

**Fig.6** Crystal structure illustrations of the supercells for YXe in (a) CsCl and (b) *Pbam* structures, respectively. Green atoms depict Y, pink atoms present Xe. (c) The total and (d) partial DOS of YXe in CsCl and *Pbam* structures, respectively.

**Fig.7** The phonon dispersion relations, the phonon density of states, Eliashberg spectral function, $\alpha^2F(\omega)$, and the electron−phonon integral, $\lambda(\omega)$, for various Y-Xe compounds at selected pressures. (a) YXe-Pbam at 60 GPa (b)$Y_3Xe$-Pnnm at 75 GPa (c) $Y_3Xe$- *I*4/*mcm* at 150 GPa (d) $YXe_2$-*I*4/*mmm* at 100 GPa and (e) $YXe_3$-*P*4/*mmm* at 150 GPa. Circles indicate the phonon line width with a radius proportional to the strength.

**Table 1.** The calculated lattice parameters (in Angstroms) and atomic positions for the newly found Y-Xe compounds at selected pressure.

**Table2**. The calculated Tcs for various Y-Xe compounds at selected pressures.

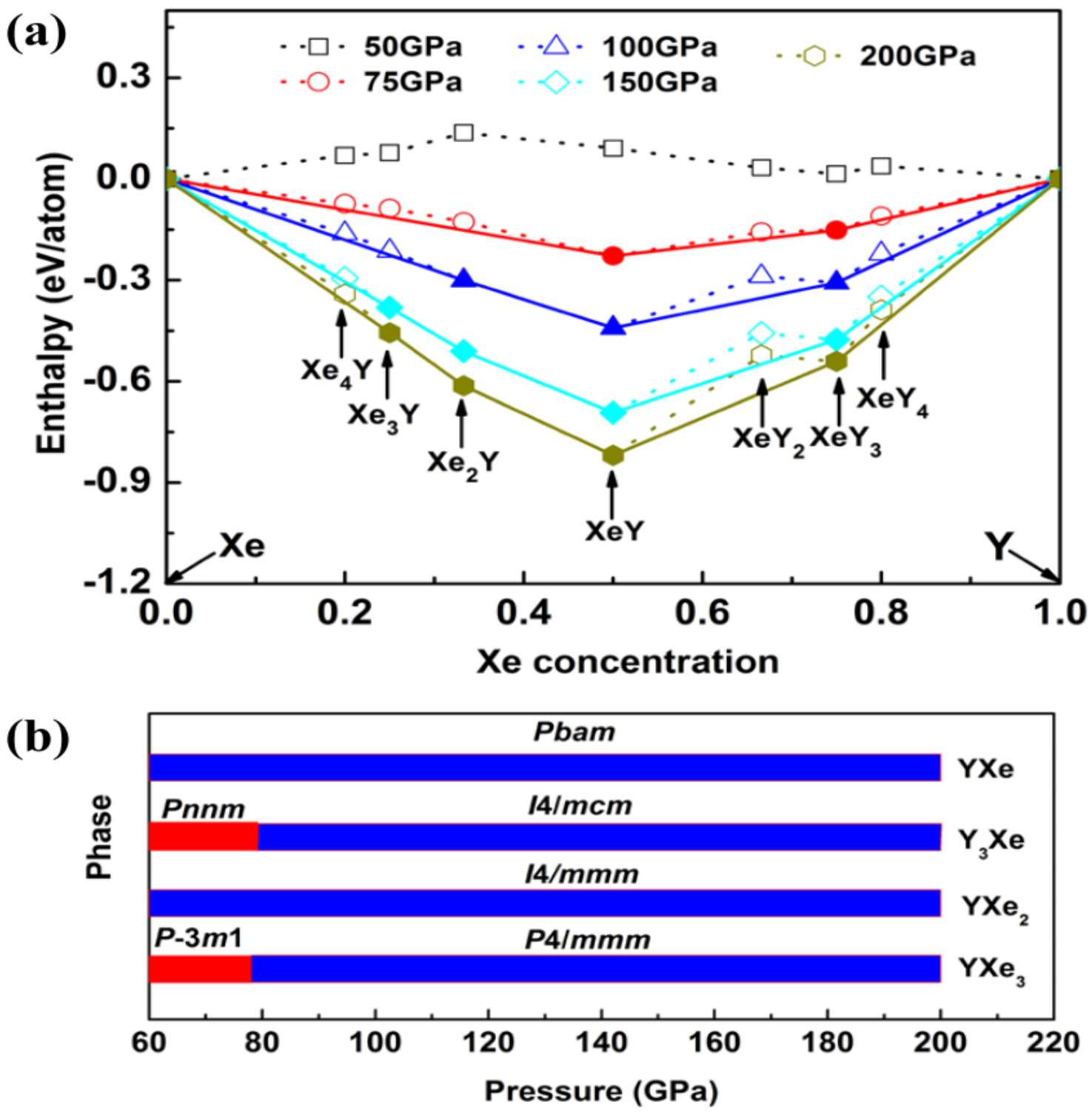

**Fig.1** (a) Ground-state enthalpies of formation for various Y−Xe compounds, relative to the elemental constituents. The dashed lines connect the data points, while the solid lines denote the convex hull. (b) The phase diagram of Y–Xe system in the pressure range from 50 to 200 GPa.

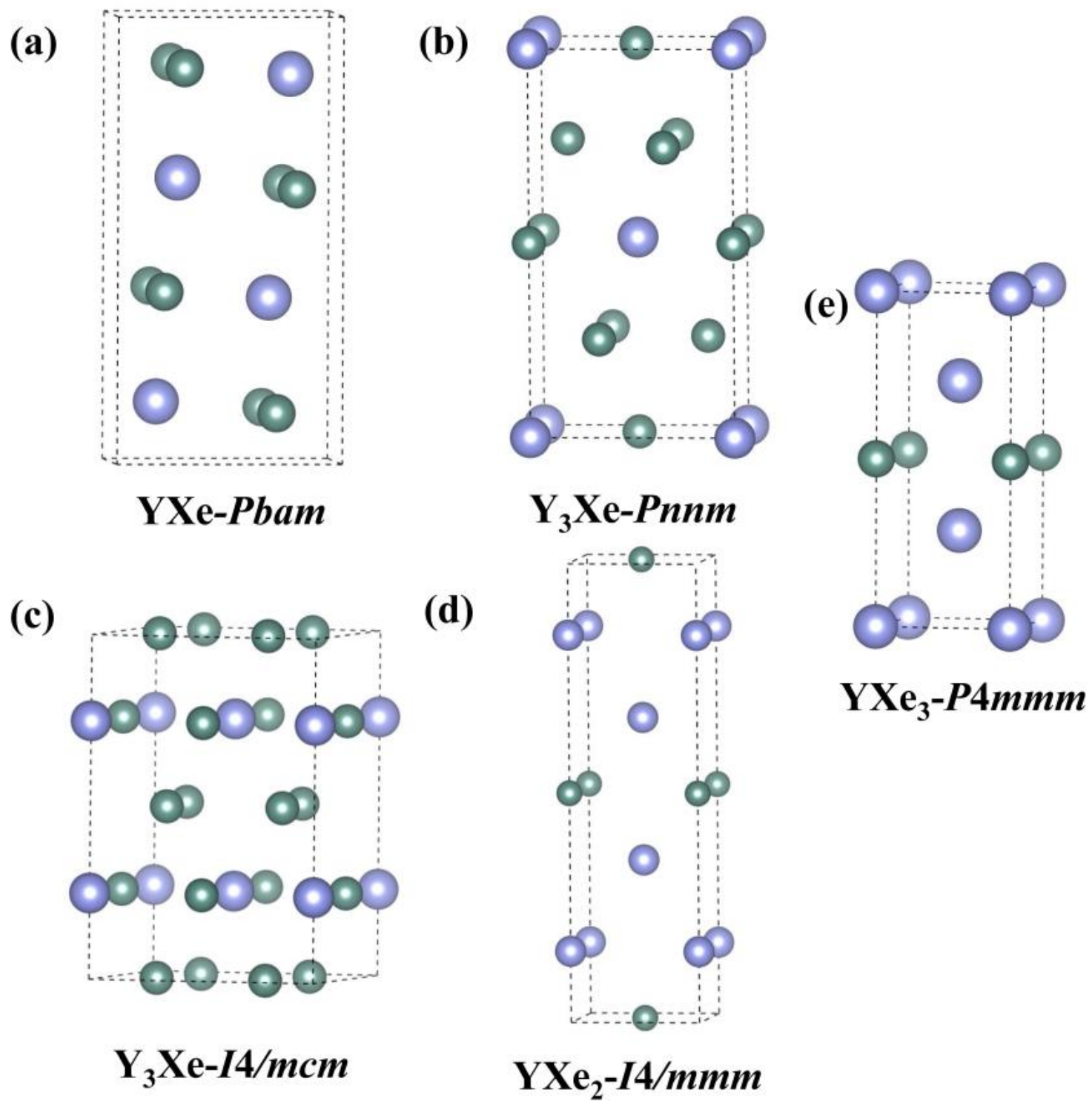

**Fig.2** Stable phases for various Y-Xe compounds at selected pressures. (a) YXe-Pbam at 75 GPa (b)$Y_3Xe$-Pnnm at 75 GPa (c) $Y_3Xe$-*I*4/*mcm* at 100 GPa (d) $YXe_2$-*I*4/*mmm* at 100 GPa and (e) $YXe_3$-*P*4/*mmm* at 150 GPa. Green atoms depict Y, pink atoms present Xe.

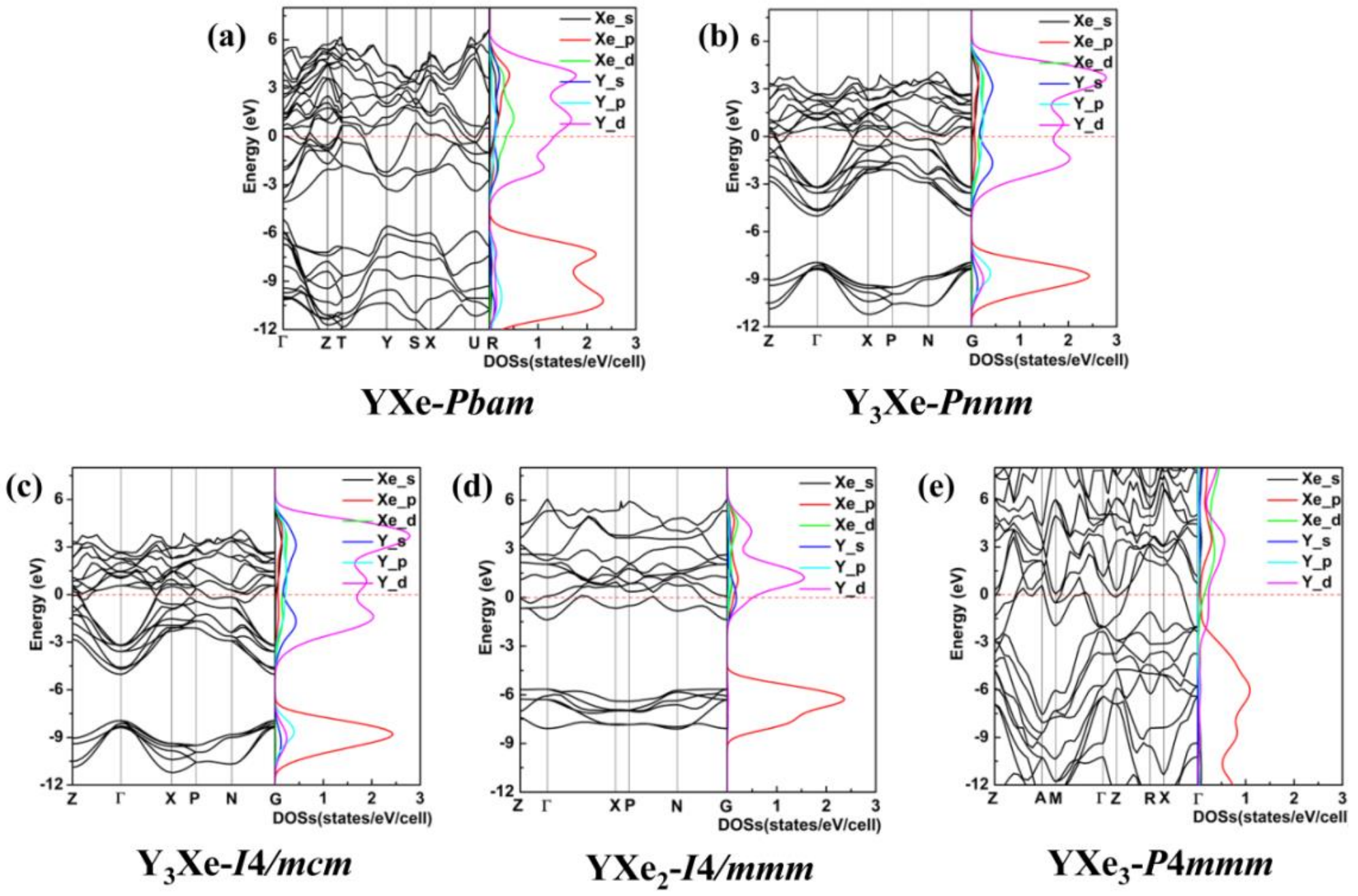

**Fig.3** Calculated electronic band structures and partial DOS for various Y−Xe compounds. (a) YXe-Pbam at 75 GPa, (b)$Y_3Xe$-Pnnm at 75 GPa, (c) $Y_3Xe$-*I*4/*mcm* at

100 GPa (d) $YXe_2$-*I*4/*mmm* at 100 GPa and (e) $YXe_3$-*P*4/*mmm* at 150 GPa.

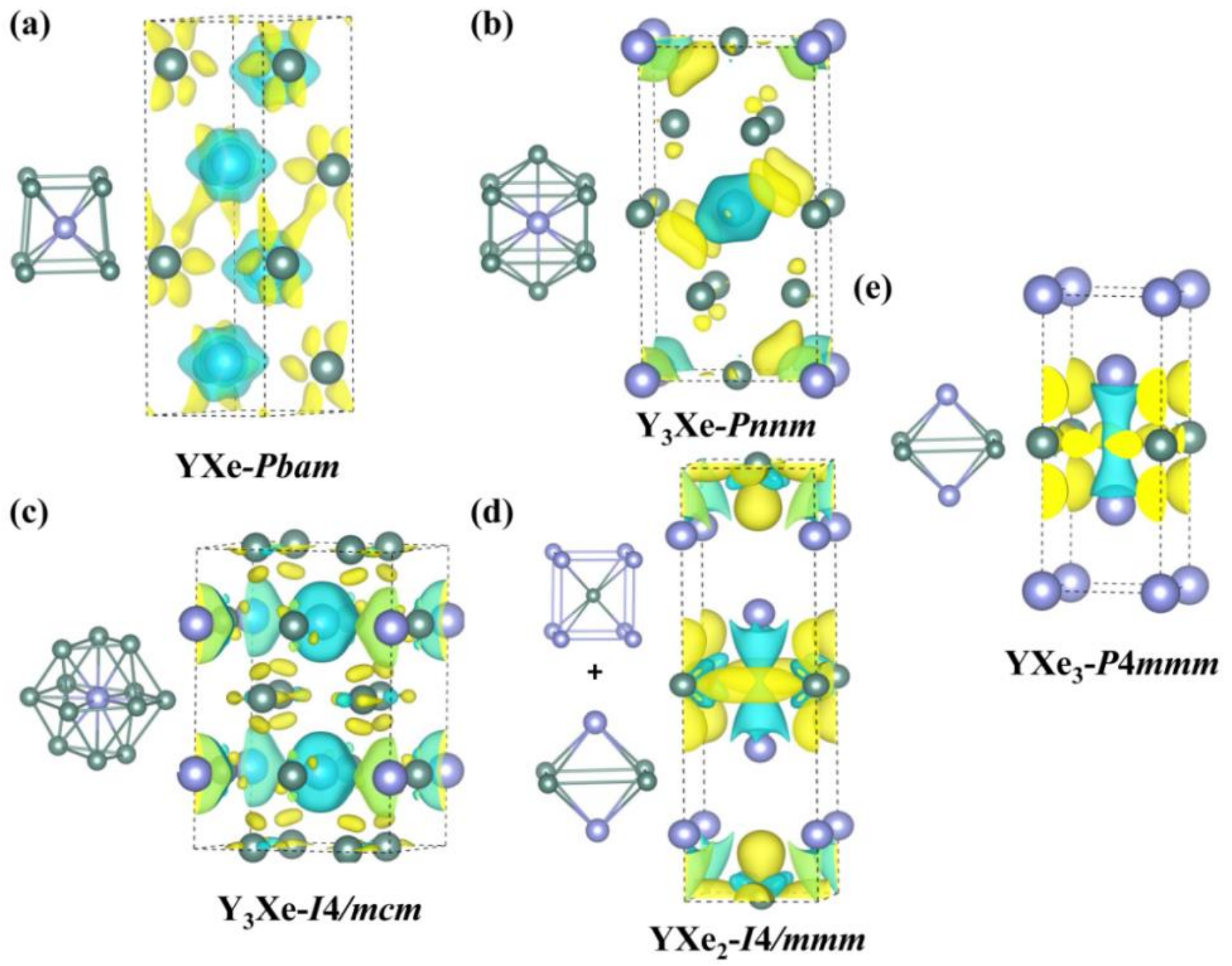

**Fig.4** Calculated charge difference density map of various Y-Xe compounds (a) YXe-Pbam at 75 GPa, (b)$Y_3Xe$-Pnnm at 75 GPa, (c) $Y_3Xe$- *I*4/*mcm* at 100 GPa (d) $YXe_2$-*I*4/*mmm* at 100 GPa and (e) $YXe_3$-*P*4/*mmm* at 150 GPa. Green atoms depict Y, pink atoms present Xe. Cyan and yellow colors represent losing and gaining electrons, respectively. The typical structure units of various Y-Xe compounds are illustrated at the left hand side of the corresponding structures.

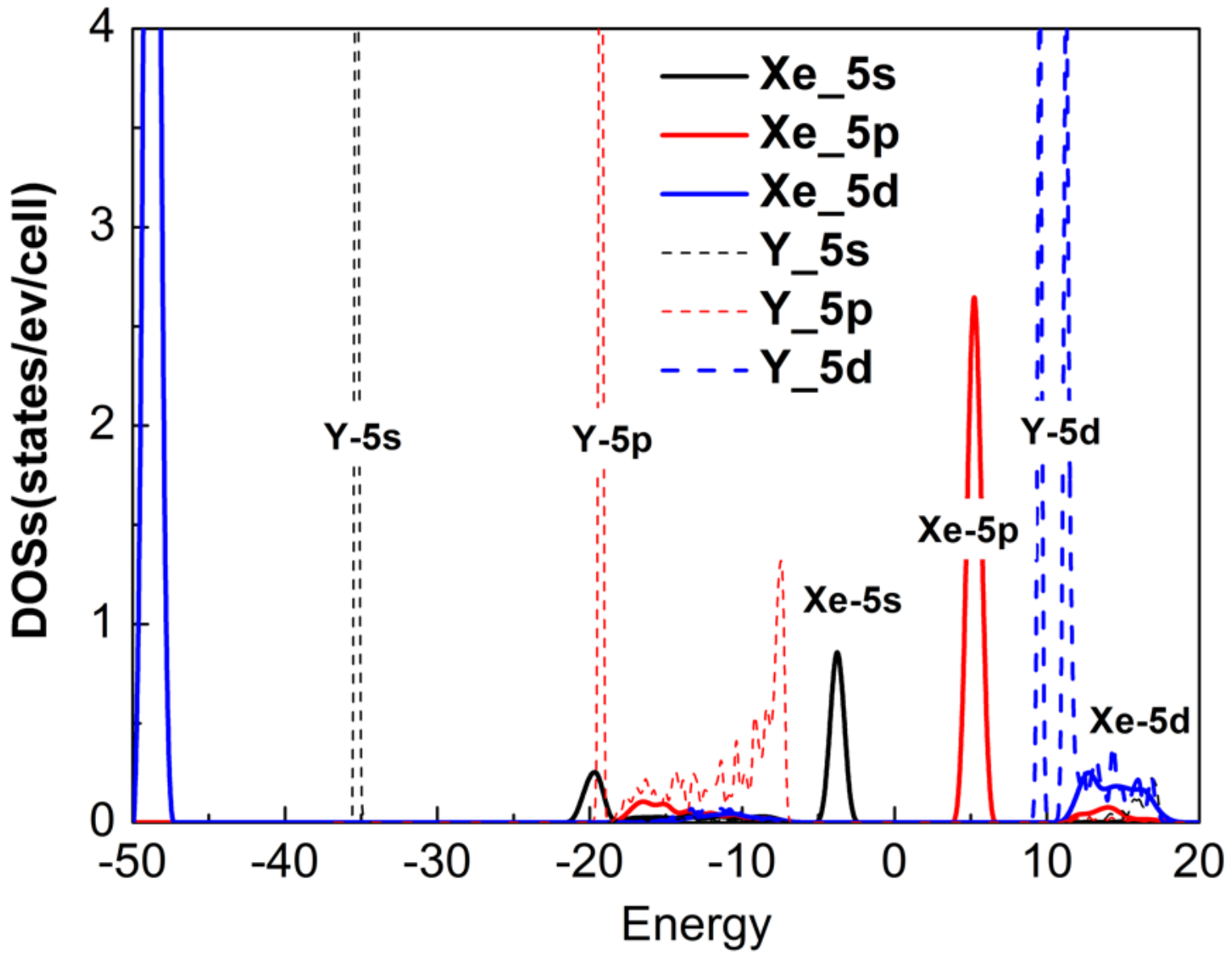

**Fig.5** Simulated partial DOSs of Xe and Y atoms at 75 GPa using the "He matrix method". The pressure effect is simulated by placing elements in a face-centered-cubic (FCC) He matrix. An FCC 3×3×3 supercell with 108 atoms was used, in which one He is replaced by the tested atoms.

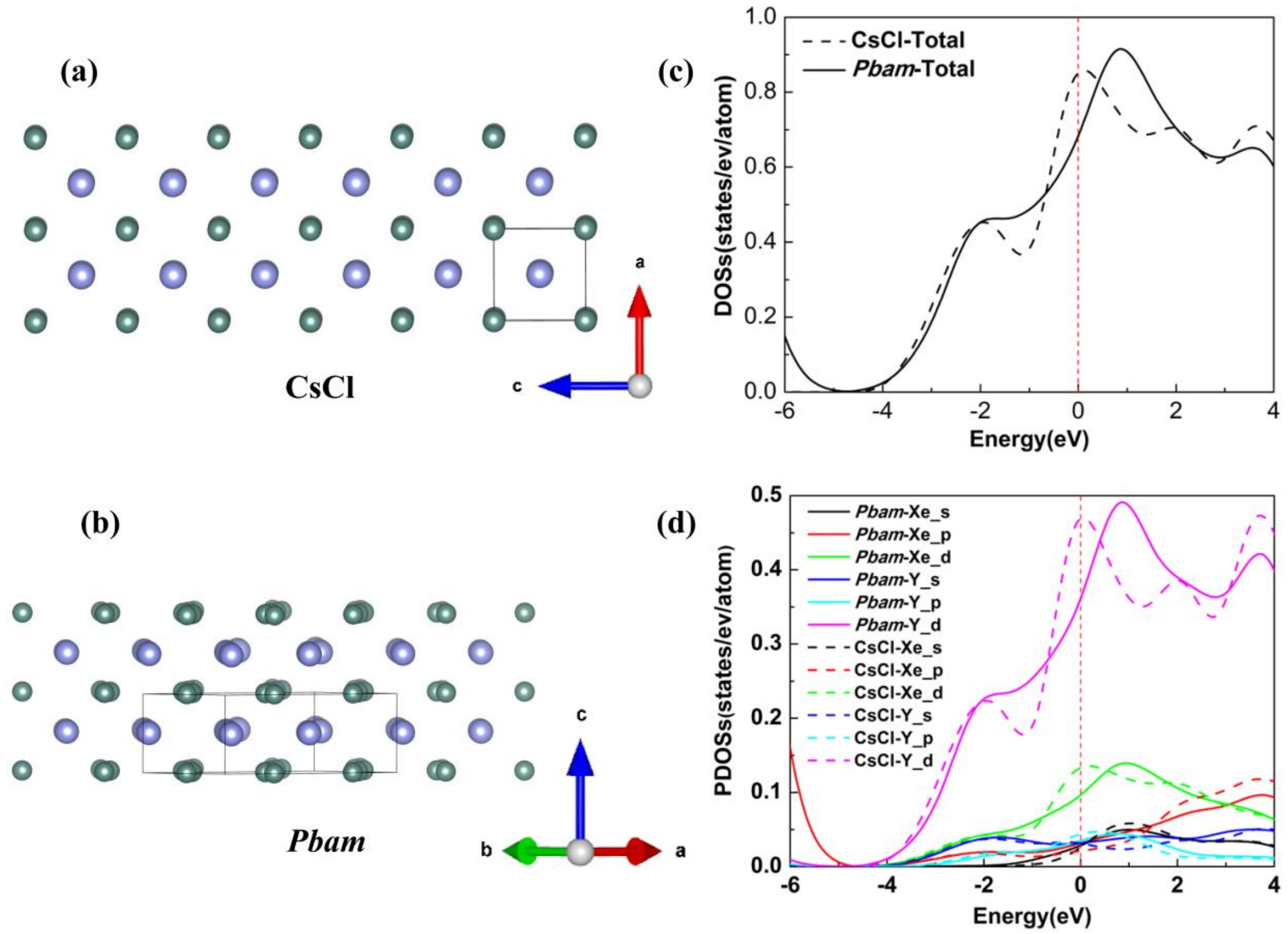

**Fig.6** Crystal structure illustrations of the supercells for YXe in (a) CsCl and (b) *Pbam* structures, respectively. Green atoms depict Y, pink atoms present Xe. (c) The total and (d) partial DOS of YXe in CsCl and *Pbam* structures, respectively.

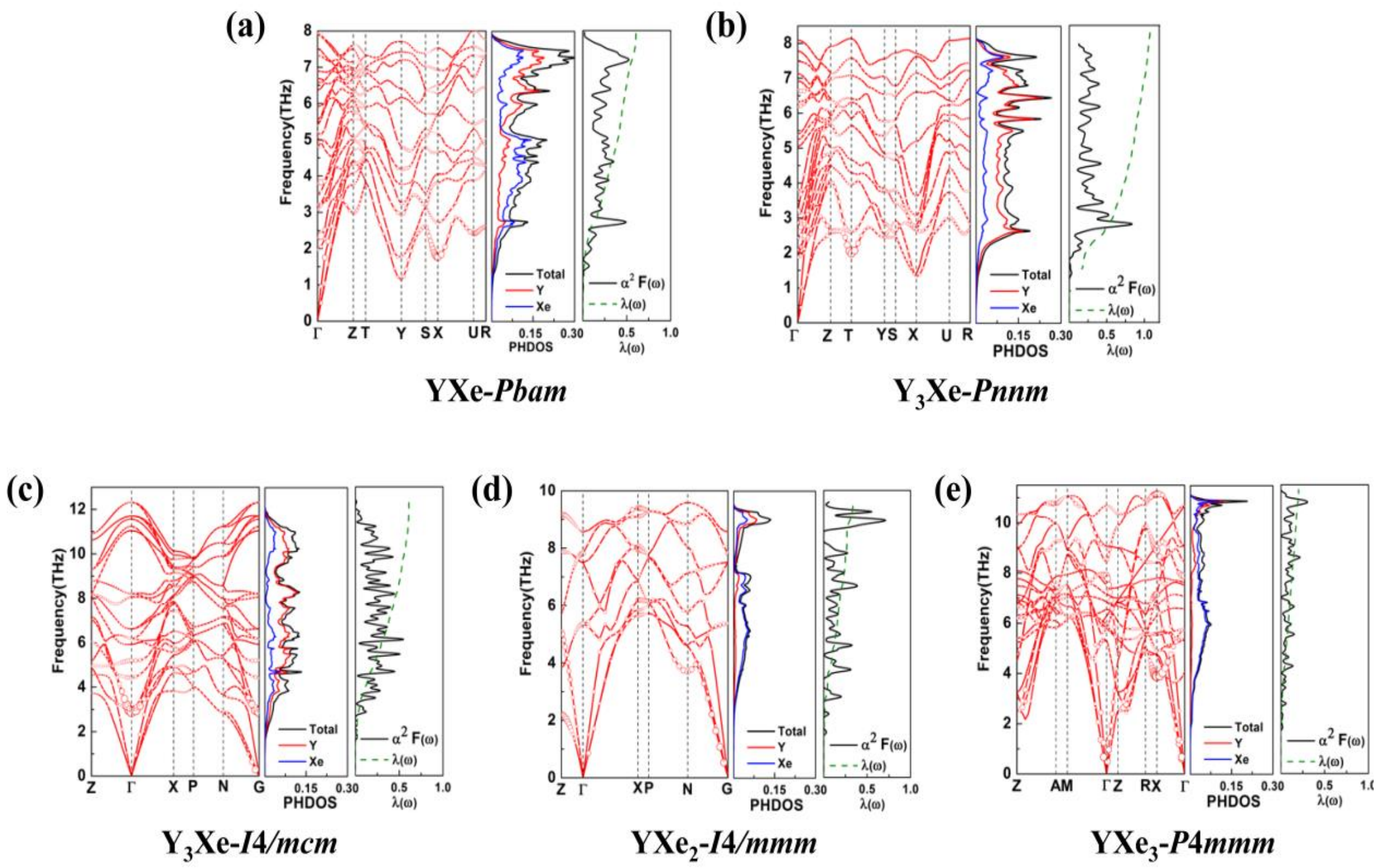

**Fig.7** The phonon dispersion relations, the phonon density of states, Eliashberg spectral function ($\alpha^2F(\omega)$) and the electron−phonon integral ($\lambda(\omega)$) for various Y-Xe compounds at selected pressures. (a) YXe-Pbam at 60 GPa (b)$Y_3$Xe-Pnnm at 75 GPa (c) $Y_3$Xe- *I*4/*mcm* at 150 GPa (d) $YXe_2$-*I*4/*mmm* at 100 GPa and (e) $YXe_3$-*P*4/*mmm* at 150 GPa. Circles indicate the phonon linewidth with a radius proportional to the strength.

**Table 1.** The calculated lattice parameters (in Angstroms) and atomic positions for the newly found Y-Xe compounds at selected pressure.

| | Space group | Pressure (GPa) | lattice parameters | Atom | Site | Wyckoff positions | | |
|---|---|---|---|---|---|---|---|---|
| YXe | *Pbam* | 75 | *a*= 9.3242 | Y | 4g | 0.3856 | 0.2070 | 0.0000 |
| | | | *b*= 4.7151 | Xe | 4h | 0.3659 | 0.7040 | 0.5000 |
| | | | *c*= 3.114 | | | | | |
| | | | $\alpha=\beta=\gamma$= 90.0 | | | | | |
| $Y_3Xe$ | *Pnnm* | 75 | *a*=*b*=5.5028 | Y | 4g | 0.3395 | 0.2531 | 0.0000 |
| | | | *c*=4.2292 | Y | 2c | 0.0000 | 0.5000 | 0.0000 |
| | | | $\alpha=\beta=\gamma$= 90.0 | Xe | 2a | 0.5000 | 0.5000 | -0.5000 |
| | *I4/mcm* | 100 | *a*=*b*= 5.463 | Y | 8h | -0.2229 | 0.7229 | 0.0000 |
| | | | *c*= 8.0713 | Y | 4b | -0.5000 | 0.0000 | -0.2500 |
| | | | $\alpha=\beta=\gamma$= 90.0 | Xe | 4a | 0.0000 | 0.0000 | 0.2500 |
| $YXe_2$ | *I4/mmm* | 100 | *a*=*b*= 5.463 | Y | 2b | 0.0000 | 0.0000 | 0.5000 |
| | | | *c*= 8.0713 | Xe | 4e | 0.0000 | 0.0000 | 0.8447 |
| | | | $\alpha=\beta=\gamma$= 90.0 | | | | | |
| $YXe_3$ | *P4/mmm* | 150 | *a*=*b*= 2.9515 | Y | 1b | 0.0000 | 0.0000 | 0.5000 |
| | | | *c*=7.18048.07 | Xe | 2h | 0.5000 | 0.5000 | 0.2768 |
| | | | $\alpha=\beta=\gamma$=90.0 | Xe | 1a | 0.0000 | 0.0000 | 0.0000 |

**Table2.** The calculated $T_C$s for various Y-Xe compounds at selected pressures.

| pressure / phase | 60 GPa | 65 GPa | 75 GPa | 100 GPa | 150 GPa | 200 GPa |
|---|---|---|---|---|---|---|
| *Pbam*-YXe | 4.4 K | - | - | 3.6 K | 3.9 K | 3.6 K |
| *Pnnm*-$Y_3$Xe | - | 7.0 K | 10.8 K | - | - | - |
| *I4/mcm*-$Y_3$Xe | - | - | - | 5.7 K | 6.5 K | 0.4 K |
| *I4/mmm*-$YXe_2$ | - | - | - | 0.3 K | - | 0.03K |
| *P4mmm*-$YXe_3$ | - | - | - | - | 0.0 K | 0.0 K |